\documentclass[conference]{IEEEtran}
\IEEEoverridecommandlockouts

\usepackage{amsmath,amssymb,amsfonts}
\usepackage{algorithmic}
\usepackage{graphicx}
\usepackage{textcomp}
\usepackage{xcolor}
\usepackage{float}
\def\BibTeX{{\rm B\kern-.05em{\sc i\kern-.025em b}\kern-.08em
    T\kern-.1667em\lower.7ex\hbox{E}\kern-.125emX}}
\usepackage[backend=biber,style=ieee]{biblatex}
\begin{document}

\title{\LARGE{Semantic Extraction Framework for Goal-Oriented V2X Communication Supporting Safe Intersection Passage}}

\author{\IEEEauthorblockN{Haoran Yang, Shan Lu, Takaya Yamazato}
\IEEEauthorblockA{\textit{Nagoya University} 
Nagoya, Japan \\
yang.haoran.g1@s.mail.nagoya-u.ac.jp, shan.lu.jp@ieee.org, yamazato@ieee.org}
}

\maketitle

\begin{abstract}
This paper proposes a transmitter-side semantic extraction framework for goal-oriented Vehicle-to-Everything (V2X) communication supporting safe intersection passage. Instead of transmitting raw sensory data, roadside infrastructure extracts task-relevant semantic information from video and traffic-signal sequences and organizes them into compact semantic packets for low-latency driving-decision assistance. The intersection-passing problem is formulated as a set of traffic tasks under Japanese traffic regulations, and the proposed pipeline generates perception and intent semantics through object detection, object tracking, region-of-interest (ROI)-based semantic extraction, and recommended-action generation. The framework is implemented and evaluated on a self-constructed, CARLA-based single-vehicle dataset collected under clear weather conditions. Experimental results show high recommended-action accuracy within the tested task set, while the mean processing latency remains below the frame interval of the 15~fps input video.
\end{abstract}


\section{Introduction}

With the rapid development of Intelligent Transportation Systems (ITS) and Vehicle-to-Everything (V2X) communication, information exchange among vehicles, roadside infrastructure, and traffic-management systems has become essential for improving traffic safety and efficiency. In particular, cooperative perception enables vehicles to obtain traffic information beyond the range of their onboard sensors, which is especially valuable at intersections, where traffic participants, signal phases, lane directions, and occluded regions jointly determine vehicle behavior. Since roadside infrastructure can be mounted at a higher, more stable position with a wider observation range and less occlusion, infrastructure-assisted V2X communication is a promising approach for supporting safe vehicle passage at intersections.
The integration of collective perception with ITS smart infrastructure
provides an architectural basis for roadside sensing and hybrid RF/VLC
information sharing~\cite{yamazato2026ictplatform}.
This perspective motivates the present focus on extracting decision-relevant information at the roadside.

Existing cooperative perception systems are commonly categorized by the type of shared information into raw sensor data, intermediate feature data, and perception-result data \cite{bai2024survey}, corresponding to early, intermediate, and late fusion, respectively, and have been widely studied \cite{chen2019cooper,wang2020v2vnet,xu2022v2xvit}. However, transmitting raw data or high-dimensional features incurs considerable communication and computational overhead, leading to unacceptable latency in bandwidth-constrained and delay-sensitive V2X scenarios. 

Visible light communication (VLC) offers a complementary channel for V2X information sharing. Nakano et al.~\cite{nakano2026vlccp} experimentally investigated VLC-based collective perception, while Soga et al.~\cite{soga2025, soga2026latency} evaluated vehicular VLC latency using multiple LED transmitters and an event-based camera. These studies motivate a complementary question addressed here: how to convert roadside observations into compact information for intersection-driving assistance.

Semantic communication offers a promising alternative, transmitting only task-relevant information instead of reconstructing all source data and preserving the meaning useful for a specific task rather than bit-level accuracy. Recent vehicular studies have explored task-oriented or goal-oriented semantic communication to reduce overhead while maintaining task performance, including extracting and protecting task-important regions for joint vehicle and license-plate recognition \cite{wu2026goal}, applying semantic-aware masking and discrete codebook design for robustness under low signal-to-noise-ratio conditions \cite{li2025robust}, and evaluating semantic reconstruction in vehicular networks by downstream task performance such as semantic-segmentation accuracy \cite{jin2025goal}.

Nevertheless, existing task-oriented V2X semantic communication studies have two major limitations for intersection-passing scenarios. First, their objectives are usually perception subtasks, such as object detection, semantic segmentation, or vehicle and license-plate recognition, rather than the final goal of safe intersection-passage assistance, which instead requires decision-level information such as whether to stop, proceed straight, turn, or prepare for a potential conflict. Second, the transmitted carriers, often cropped image regions, reconstructed features, or neural-network representations, still require additional decoding or inference at the receiver, which is undesirable for latency, reliability, and resource-sensitive onboard devices.

\begin{figure}
    \centering
    \includegraphics[width=0.7\linewidth]{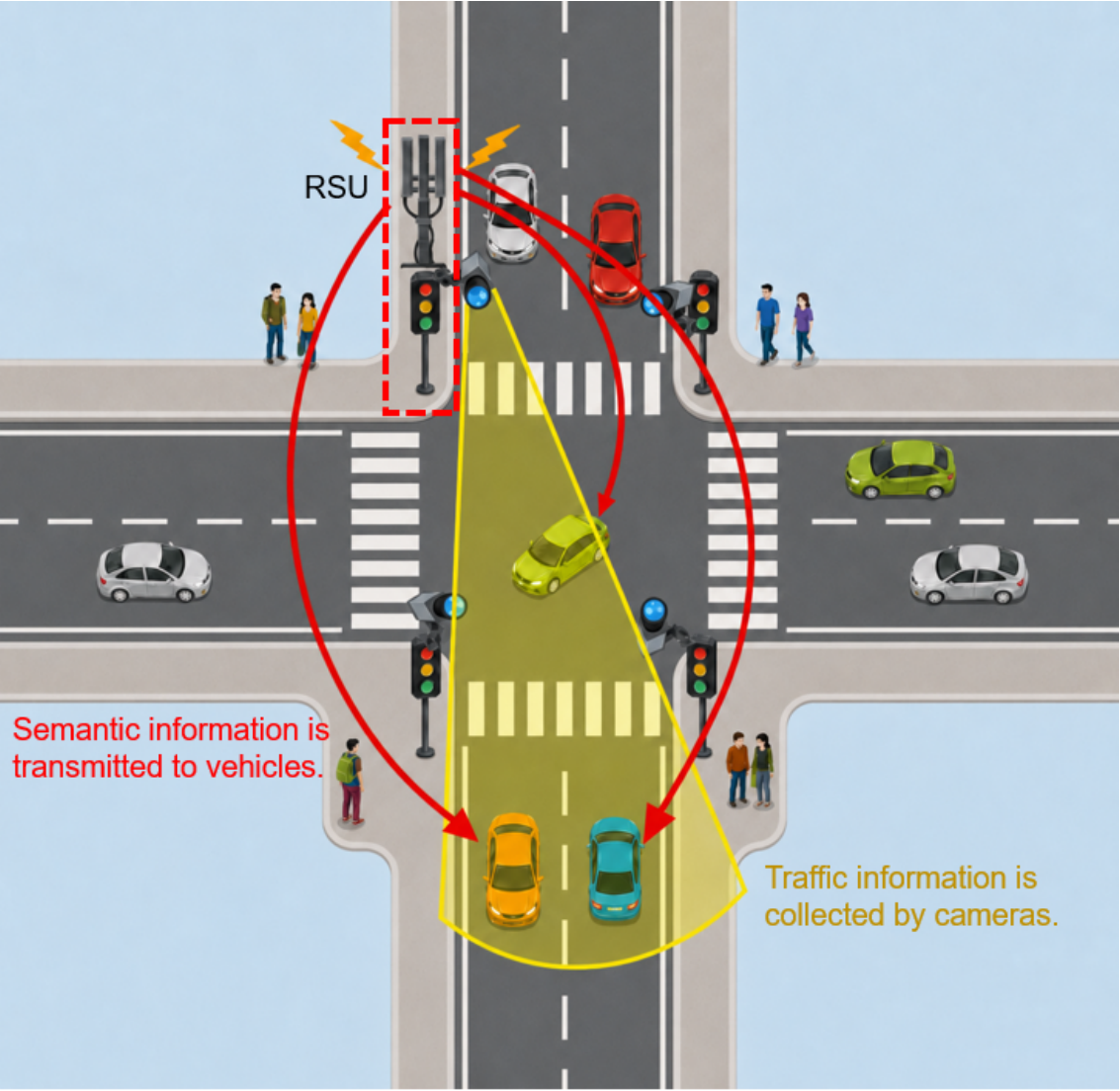}
    \caption{Vehicle-infrastructure cooperative architecture}
    \label{fig:cooperative_architecture}
\end{figure}

To address these problems, this paper focuses on how to extract and transmit the minimum decision-relevant semantic information from roadside perception so that vehicles can complete safe intersection-passing tasks under communication and latency constraints. Specifically, we propose a goal-oriented V2X semantic communication system that treats safe intersection passage as the system-level goal and formulates intersection-passing scenarios as traffic tasks. Exploiting the observation height and wide coverage of roadside infrastructure, the transmitter extracts traffic-perception and traffic-intent semantic information, including traffic-signal state, remaining signal time, region information, priority, and recommended driving action, and packages them into compact, low-dimensional semantic packets. These packets provide vehicles with decision-relevant information for real-time driving-decision assistance. In this way, the proposed system shifts the communication objective from scene reconstruction or perception-result sharing to task completion, thereby potentially reducing redundant transmission and receiver-side computational burden, as shown in Fig.~\ref{fig:cooperative_architecture}.

The main contributions of this paper are summarized as follows:

\begin{itemize}
    \item This paper formulates safe intersection passage as a goal-oriented V2X semantic communication task and defines a traffic-task set under Japanese traffic regulations.
    \item This paper designs a transmitter-side semantic extraction framework that converts roadside video and signal sequences into compact semantic packets containing perception and intent semantic information.
    \item This paper implements the framework in CARLA-based single-vehicle scenarios and evaluates it in terms of recommended-action accuracy, task-completion rate, processing latency, and packet-size reduction.
\end{itemize}

\section{Proposed System Model}

\subsection{Traffic Task Model}
This study takes the safe intersection-passage assistance of a vehicle through an intersection as the target task of the communication system. Unlike conventional V2X communication, which aims at faithfully transmitting raw sensor data, the proposed system focuses only on delivering the information a vehicle needs for safe intersection-passage assistance. Such a task is jointly determined by the vehicle maneuver, traffic-signal state, spatial relation to predefined ROIs, interaction condition with surrounding participants, and the corresponding recommended action.

According to the level of interaction with these participants, intersection-passing tasks span four categories of increasing complexity: no interaction (ego-only), interaction with vehicles, with vulnerable road users (VRUs) such as pedestrians and cyclists, and with priority units such as ambulances and fire engines. Ego-only tasks depend only on the target vehicle and the traffic signal and are the basic building blocks of complex scenarios, whereas interaction tasks additionally require spatial-relationship reasoning, right-of-way judgment, and conflict-risk analysis. As the fundamental validation scenario, this paper first studies the ego-only task at a common four-way intersection formed by two crossing bidirectional four-lane roads, where each direction has a shared straight-and-left-turn lane and a dedicated right-turn lane. Rather than adopting a single perception task as the system goal, we therefore organize the perception results, under Japanese traffic regulations, into traffic tasks oriented toward behavior judgment, each combining the vehicle maneuver, the traffic-signal state, the spatial region, the interaction condition, and the recommended action, with the maneuver and recommended-action sets defined as
\begin{IEEEeqnarray}{rCl}
\mathcal{M} &=& \{\text{Straight},\ \text{Left Turn},\ \text{Right Turn}\}
\IEEEyesnumber\\
\mathcal{A} &=& \{\text{GO\_STRAIGHT\_LEFT},\ \text{GO\_RIGHT},
\nonumber\\
&& \ \text{STOP},\ \text{PREPARE}\}
\IEEEyesnumber
\end{IEEEeqnarray}

These define the intersection-passing task set $\mathcal{T} = \{T_n \mid n = 1,2,\ldots,N\}$, where $N$ is the number of predefined tasks and each task is represented as
\begin{equation}
T_n
=
\left(
m_n,
\ell_n,
r_n,
z_n,
a_n
\right),
\end{equation}

where $m_n \in \mathcal{M}$ is the maneuver, $\ell_n \in \mathcal{L}$ the traffic-signal condition, $r_n$ the spatial condition given by the ROI (region-of-interest) relation, $z_n$ the interaction condition with surrounding participants, and $a_n \in \mathcal{A}$ the desired action.

The system aims not to recover the complete sensor data but to judge and transmit the key semantic information required to complete task $T_n$, which is regarded as completed when the system correctly outputs a recommended action consistent with traffic regulations and safe intersection-passage assistance. We accordingly evaluate the system by the task-completion rate,
\begin{equation}
\eta_{\mathrm{task}}
=
\frac{N_{\mathrm{completed}}}
     {N_{\mathrm{total}}},
\end{equation}
where $N_{\mathrm{completed}}$ and $N_{\mathrm{total}}$ are the numbers of correctly completed tasks and of all tasks in the set.

The ego-only recommended actions are the four in $\mathcal{A}$ : proceeding from the straight-and-left-turn lane (\text{GO\_STRAIGHT\_LEFT}), proceeding from the right-turn lane (\text{GO\_RIGHT}), stopping before the stop line (\text{STOP}), and preparing to start as the red light is about to end (\text{PREPARE}). Each task is organized by maneuver (straight, left-turn, right-turn) and signal state (green, yellow, red), and the intersection is divided into regions of interest (ROIs), such as the straight-and-left-turn lane, the right-turn lane, and the buffer zone for lane changing. Table~\ref{tab:Traffic Task Classification} lists the basic ego-only tasks considered in this paper.

\begin{table*}[t]
\centering
\caption{Traffic Task Classification}
\vspace{0.5em}
\footnotesize
\begin{tabular}{ccccc}
\hline
\textbf{Task No.} & \textbf{Interaction level} & \textbf{Maneuver} & \textbf{Signal} & \textbf{Description} \\
\hline
T1 & Ego-only & Straight & Green & clear + correct lane \\
T2 & Ego-only & Straight & Yellow & can stop safely \\
T3 & Ego-only & Straight & Yellow & cannot stop safely + clear \\
T4 & Ego-only & Straight & Red & Red light \\
T5 & Ego-only & Left turn & Green & clear + correct lane \\
T6 & Ego-only & Left turn & Yellow & can stop safely \\
T7 & Ego-only & Left turn & Yellow & cannot stop safely + clear \\
T8 & Ego-only & Left turn & Red & Red light \\
T9 & Ego-only & Right turn & Green & clear + correct lane \\
T10 & Ego-only & Right turn & Yellow & can stop safely \\
T11 & Ego-only & Right turn & Yellow & cannot stop safely + clear \\
T12 & Ego-only & Right turn & Red & Red light \\
\hline
\end{tabular}
\label{tab:Traffic Task Classification}
\end{table*}

For the $i$-th frame, the system extracts traffic semantic information $\mathcal{S}_i$ and infers a recommended action for each object. The task-oriented decision process is formulated as
\begin{equation}
a_{i,j}
=
f_{\mathrm{act}}
\left(
\iota_{i,j},
c_{i,j},
\boldsymbol{\gamma}_{i,j},
\ell_i,
\tau_i^{\mathrm{sig}}
\right),
\end{equation}
where $a_{i,j}$ is the recommended action for the $j$-th tracked object in the $i$-th frame, $\iota_{i,j}$ its track ID, $c_{i,j}$ its class, and $\boldsymbol{\gamma}_{i,j}$ its ROI relation. The transmitter-side semantic processing thus converts the raw sensing input $\mathcal{I}_i$ into task-relevant semantic information that let the receiver obtain $a_{i,j}$ with low latency and communication overhead.

\subsection{Overview of Our Proposed Communication System}
\begin{figure*}[t]
    \centering
    \includegraphics[width=0.9\linewidth]{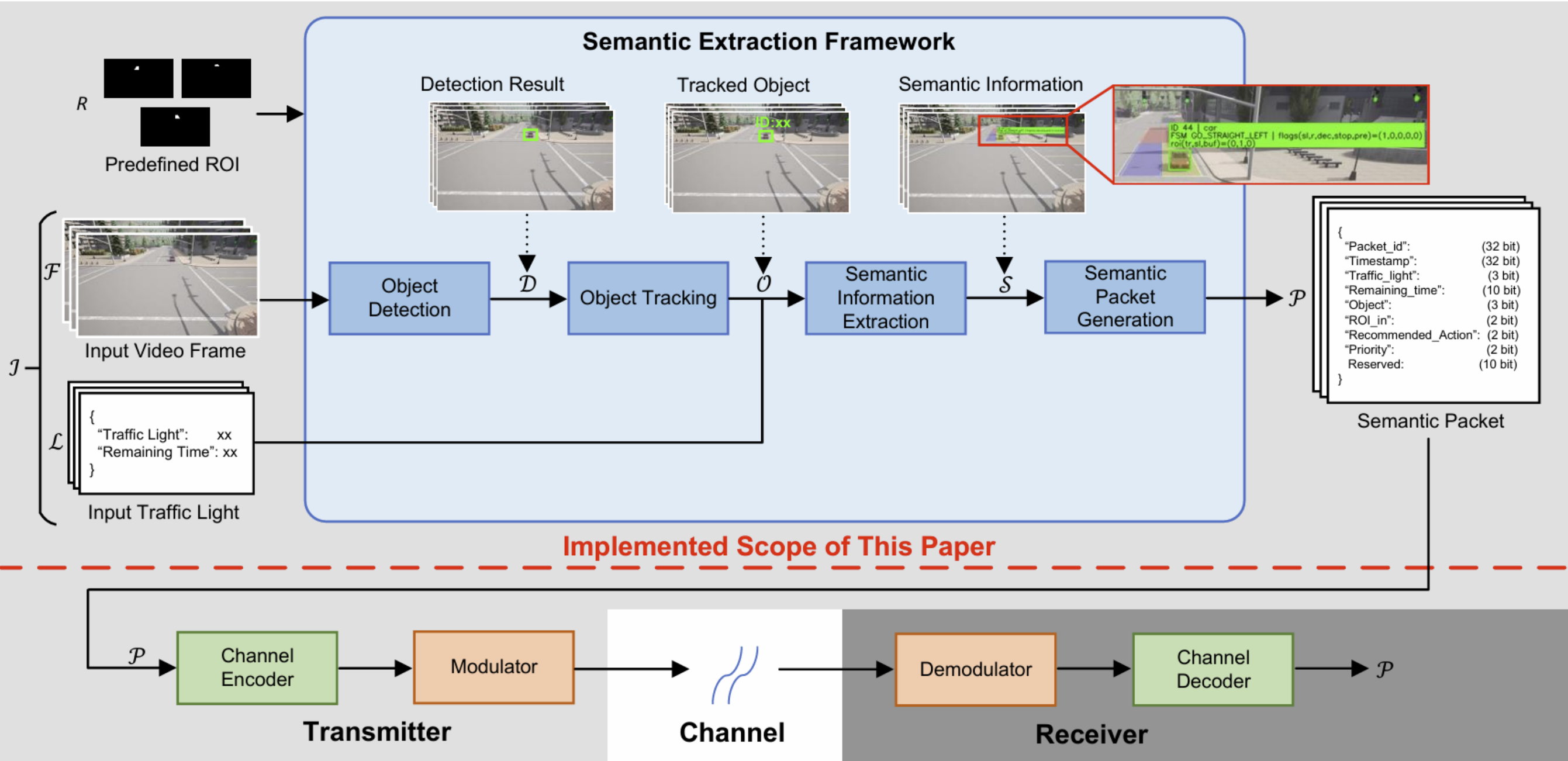}
    \caption{System model (The region above the red dashed line represents the implemented scope of this paper)}
    \label{fig:system_model}
\end{figure*}

The proposed goal-oriented V2X semantic communication architecture consists of roadside infrastructure acting as the transmitter and a vehicle acting as the receiver, as shown in Fig.~\ref{fig:system_model}. Although the overall architecture includes both transmission and reception, this paper implements and evaluates the transmitter-side semantic extraction and packet-generation pipeline as a first step toward realizing the complete system.

At the physical level, the roadside infrastructure comprises traffic signals, a roadside camera, and a control terminal. The traffic signals perform conventional traffic control at the intersection. The camera, mounted at the height of the signals and oriented toward the opposite lanes, captures real-time video covering the intersection and the oncoming area. The control terminal synchronously acquires the signal states and the video, performs semantic processing on both, and generates semantic packets for transmission to the vehicle. The system thus provides vehicles with task-relevant traffic semantic information for safe passage rather than raw sensory data.

The raw traffic-sensing data consist of a video frame sequence and a traffic-signal sequence. The video stream is organized by frame indices, and each frame is associated with a capture timestamp. The video sequence is represented as
\begin{equation}
\mathcal{V}
=
\left\{
\left(\mathcal{F}_i, \tau_i^{\mathrm{cap}}\right)
\mid
i = 0,1,\ldots,N_{\mathrm{v}}-1
\right\},
\end{equation}
where $\mathcal{F}_i$ denotes the $i$-th video frame, $\tau_i^{\mathrm{cap}}$ its capture timestamp, and $N_{\mathrm{v}}$ the total number of frames. 

The traffic-signal sequence is represented as a set of signal phases,
\begin{equation}
\mathcal{G}
=
\left\{
\left(\ell_g, \beta_g\right)
\mid
g = 1,2,\ldots,G
\right\},
\end{equation}
where $\ell_g$ is the state of the $g$-th phase and $\beta_g$ its switching time. The signal state belongs to $\mathcal{L} = \{\mathrm{red},\,\mathrm{yellow},\,\mathrm{green}\}$. The remaining signal time, denoted by $\tau_i^{\mathrm{sig}}$, is obtained from the synchronized signal sequence and represents the time interval between the current frame timestamp and the next signal transition.

After time synchronization, the raw traffic-sensing input for the $i$-th frame is defined as
\begin{equation}
\mathcal{I}_i
=
\left(
\mathcal{F}_i,
\ell_i,
\tau_i^{\mathrm{sig}},
\tau_i^{\mathrm{cap}}
\right).
\end{equation}

The overall transmitter-side processing flow is then expressed as
\begin{equation}
\mathcal{I}_i
\rightarrow
\mathcal{D}_i
\rightarrow
\mathcal{O}_i
\rightarrow
\mathcal{S}_i
\rightarrow
\mathcal{P}_i,
\end{equation}
where $\mathcal{D}_i$ denotes the detection result set, $\mathcal{O}_i$ the tracked object set, $\mathcal{S}_i$ the extracted semantic information, and $\mathcal{P}_i$ the generated semantic packet.

\subsection{Transmitter-Side Semantic Processing}
The transmitter-side semantic processing consists of object detection, object tracking, and traffic semantic-information extraction. These modules transform the raw traffic-sensing input into structured semantic information suitable for packet generation.

\subsubsection{Object Detection}
The object-detection module identifies traffic participants in each video frame and outputs their classes and bounding boxes. For the $i$-th frame,
\begin{equation}
\left(
\mathcal{D}_i,
\tau_i^{\mathrm{det}}
\right)
=
f_{\mathrm{det}}(\mathcal{F}_i),
\quad
\mathcal{D}_i
=
\left\{
d_{i,j}
\mid
j = 1,\ldots,J_i
\right\},
\end{equation}
where $\tau_i^{\mathrm{det}}$ is the timestamp recorded after detection and $J_i$ the number of detected objects. Each detected object is represented as $d_{i,j} = (c_{i,j}, \mathbf{b}_{i,j}, p_{i,j})$, where $c_{i,j}$ is the class, $p_{i,j}$ the confidence score, and $\mathbf{b}_{i,j}$ the bounding box,
\begin{equation}
\mathbf{b}_{i,j}
=
\left(
x_{1},
y_{1},
x_{2},
y_{2},
c x,
c y
\right)_{i,j},
\end{equation}
in which $(x_1,y_1)$ and $(x_2,y_2)$ are the top-left and bottom-right coordinates and $(cx,cy)$ the center coordinate of the box. The detected class belongs to
\begin{equation}
\mathcal{C}
=
\{
\mathrm{pedestrian},
\mathrm{general\ vehicle},
\mathrm{priority\ vehicle}
\},
\end{equation}
where $\mathrm{priority\ vehicle}$ represents emergency or high-priority participants such as ambulances, fire trucks, and police vehicles. 

\subsubsection{Object Tracking}
The object-tracking module associates detected objects across consecutive frames and assigns a unique track identity to each object,
\begin{equation}
\left(
\mathcal{O}_i,
\tau_i^{\mathrm{trk}}
\right)
=
f_{\mathrm{trk}}
\left(
\mathcal{D}_{i}
\right),
\quad
\mathcal{O}_i
=
\left\{
o_{i,j}
\mid
j = 1,\ldots,K_i
\right\},
\end{equation}
where $\tau_i^{\mathrm{trk}}$ is the timestamp recorded after tracking and $K_i$ the number of tracked objects. Each tracked object is represented as $o_{i,j} = (\iota_{i,j}, c_{i,j}, \mathbf{b}_{i,j})$, where $\iota_{i,j}$ is the track ID, $c_{i,j}$ the class, and $\mathbf{b}_{i,j} = (x_1, y_1, x_2, y_2, cx, cy)_{i,j}$ the tracked bounding box with the same format as the detection box. Compared with frame-wise detection, tracking provides temporal consistency and object identity, which are necessary for extracting higher-level traffic semantic information such as object movement, ROI transition, and action intention.

\subsubsection{Traffic Semantic Information Extraction}

The semantic-extraction module converts the tracked object set and the synchronized signal information into task-relevant semantic representations. An object is not represented by a single point; instead, for each tracked object, the ROI relation is encoded by a binary vector $\boldsymbol{\gamma}_{i,j} = [\gamma_{i,j,1}, \ldots, \gamma_{i,j,Q}]$, where each entry $\gamma_{i,j,q}$ indicates whether the object is assigned to ROI $R_q$ and is jointly determined by the $y$-coordinate of the lower bounding-box boundary, $y_{2,i,j}$, the vertical extent $[y_q^{\mathrm{top}}, y_q^{\mathrm{bot}}]$ of $R_q$, and the bounding-box-ROI overlap ratio $\rho_{i,j,q}$ relative to a threshold $\rho_{\mathrm{th}}$.

The traffic-perception semantic information comprise the traffic-signal state, the remaining signal time, and the set of per-object ROI relations $\gamma_i = \{\boldsymbol{\gamma}_{i,j} \mid j = 1,\ldots,K_i\}$,
\begin{equation}
\mathcal{S}_i^{\mathrm{perc}}
=
\left(
\ell_i,
\tau_i^{\mathrm{sig}},
{\gamma}_{i}
\right).
\end{equation}

Based on the perception semantic information, the intent semantic information are obtained by inferring the recommended action for each tracked object,
\begin{equation}
\mathcal{S}_i^{\mathrm{int}}
=
\left\{
\left(\iota_{i,j}, a_{i,j}\right)
\mid
j = 1,\ldots,K_i
\right\}.
\end{equation}

The complete semantic information extracted from the $i$-th frame is
\begin{equation}
\left(
\mathcal{S}_i,
\tau_i^{\mathrm{sem}}
\right)
=
f_{\mathrm{sem}}
\left(
\mathcal{O}_i,
\ell_i,
\tau_i^{\mathrm{sig}},
{\gamma}_{i}
\right),
\quad
\mathcal{S}_i
=
\left(
\mathcal{S}_i^{\mathrm{perc}},
\mathcal{S}_i^{\mathrm{int}}
\right),
\end{equation}
where $\tau_i^{\mathrm{sem}}$ is the timestamp recorded after semantic extraction. The overall transmitter-side processing latency is 
\begin{equation}
\tau_i^{\mathrm{proc}} = \tau_i^{\mathrm{sem}} - \tau_i^{\mathrm{cap}}.
\end{equation}

\subsubsection{Semantic Packet Generation}
After semantic extraction, the transmitter generates a semantic packet that conveys compact task-relevant information instead of raw video frames,
\begin{equation}
\mathcal{P}_i
=
f_{\mathrm{pkt}}
\left(
\mathcal{S}_i,
\tau_i^{\mathrm{sem}}
\right)
=
\left(
H_i,
\tau_i^{\mathrm{sem}},
\ell_i,
\tau_i^{\mathrm{sig}},
\mathcal{E}_i
\right),
\end{equation}
where $H_i = (\mathrm{pid}_i, i, K_i)$ is the packet header consisting of the packet ID, the frame index, and the number of tracked objects, and $\mathcal{E}_i = \{e_{i,j} \mid j = 1,\ldots,K_i\}$ is the set of object-level semantic entries. Each entry is
\begin{equation}
e_{i,j}
=
\left(
\iota_{i,j},
c_{i,j},
\boldsymbol{\gamma}_{i,j},
a_{i,j}
\right).
\end{equation}
Compared with transmitting the full bounding box and image frame, an entry keeps only the object identity, class, ROI relation, and recommended action, all directly related to the intersection-passing task and suitable for low-rate V2X communication.

Since the packet is used for real-time traffic assistance, the semantic information must remain valid when it is received and decoded, which imposes the end-to-end latency constraint
\begin{equation}
\tau_i^{\mathrm{proc}}
+
\tau_i^{\mathrm{tx}}
+
\tau_i^{\mathrm{dec}}
\leq
\tau_i^{\mathrm{valid}},
\end{equation}
where $\tau_i^{\mathrm{tx}}$ is the transmission latency, $\tau_i^{\mathrm{dec}}$ the receiver-side decoding latency, and $\tau_i^{\mathrm{valid}}$ the validity time of the semantic information.

\section{Experiment}

\subsection{Experiment Setup}

The entire transmitter-side processing pipeline is implemented in Python and executed on a laptop with an NVIDIA RTX 4060 Laptop GPU, emulating the computational resources of a roadside control terminal. Object detection adopts the YOLO11s model, and object tracking employs the BoT-SORT algorithm to maintain consistent track identities across frames. The intent semantic information is generated by a finite-state machine (FSM).

This study focuses on simplified single-vehicle scenarios as a first step, where a single vehicle enters the intersection and performs one of the defined traffic tasks. These tasks are validated through CARLA simulations. The relevant dataset parameters are listed in Table~\ref{tab:experimental_setup}.

\begin{table}[H]
\centering
\caption{Dataset Parameter}
\vspace{0.5em}
\begin{tabular}{cc}
\hline
\textbf{Parameter} & \textbf{Value} \\
\hline
Camera Height & 5.5 m \\
Camera Pitch Angle & $90^\circ$ \\
Camera Field of View (FOV) & $90^\circ$ downward \\
Image Resolution & $1280 \times 720$ pixels \\
Frame Rate & 15 fps \\
Number of Frames & 2250 (covering all traffic tasks) \\
\hline
\end{tabular}
\label{tab:experimental_setup}
\end{table}

\subsection{Simulation Results}

\subsubsection{Accuracy of Intent Extraction}

To evaluate the accuracy of the extracted intent semantic information, the recommended action predicted by the transmitter for the ego vehicle is compared frame by frame with the ground-truth action of the corresponding traffic task. The results within the tested single-vehicle task set are reported in Table~\ref{tab:recommended_action_summary}. Since the dataset is generated by CARLA, the exact vehicle coordinates and distance to the intersection are available as ground truth, whereas the vehicle-to-ROI relation in the top-down camera view must be inferred under perspective distortion. The few remaining errors mainly occur in transition frames near ROI boundaries, where this relation is momentarily misjudged. In our ROI design, these boundary-transition regions are located far from the intersection conflict area, and thus have limited influence on safety-critical actions such as stopping or preparing to start.

\begin{figure}[hbpt]
    \centering
    \includegraphics[width=0.8\linewidth]{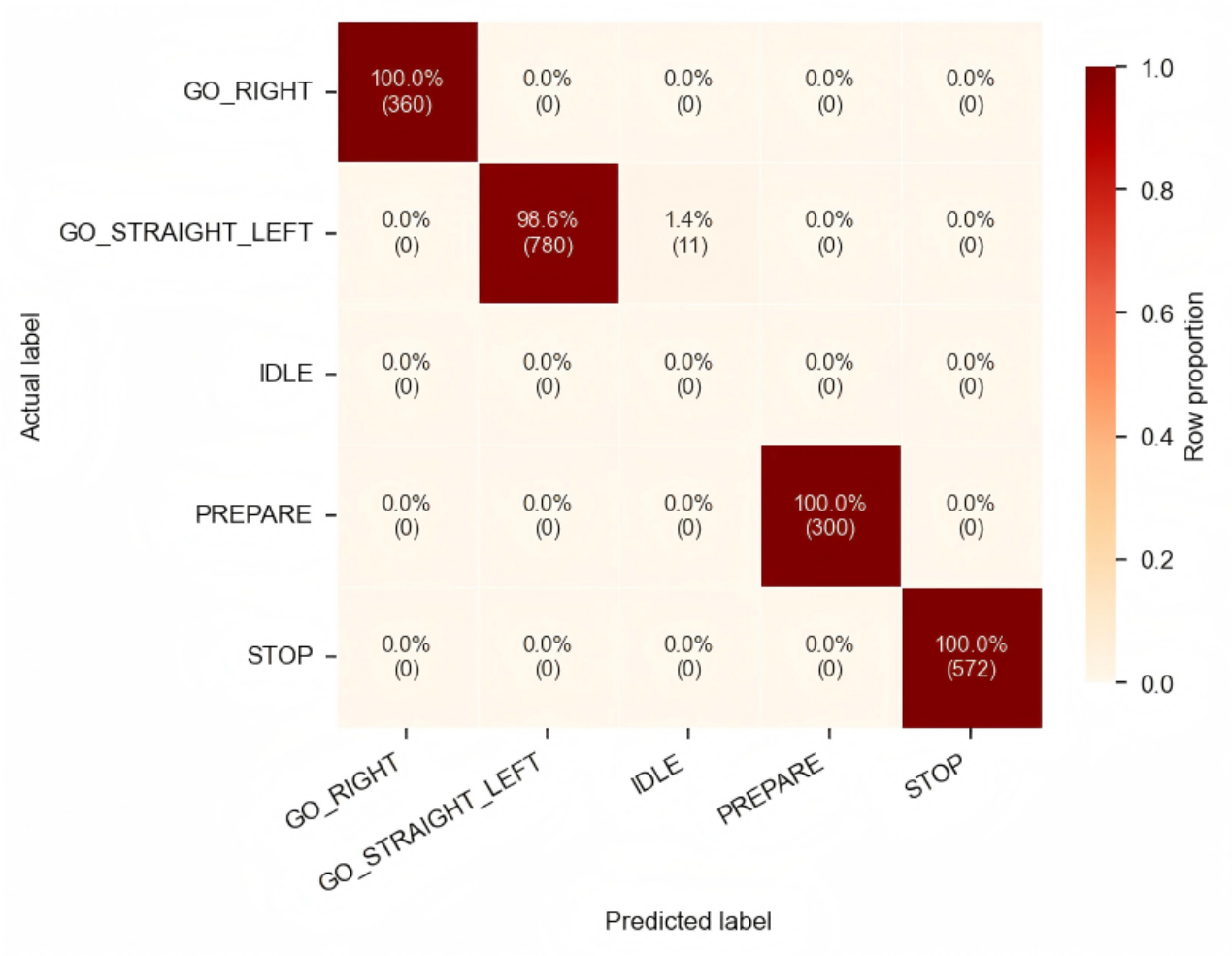}
    \caption{Confusion Matrix of recommended action prediction across all traffic tasks}
    \label{fig:action_confusion}
\end{figure}

Fig.~\ref{fig:action_confusion} shows the confusion matrix of the recommended-action prediction aggregated over all traffic tasks, including the initial IDLE state of the FSM. For the same reason, the small number of errors is concentrated in the transition from the initial IDLE state to the adjacent action GO\_STRAIGHT\_LEFT, whereas no errors are observed for STOP and PREPARE within the tested single-vehicle task set.

\begin{table}[h]
\centering
\caption{Recommended-action prediction accuracy by traffic task.}
\label{tab:recommended_action_summary}
\footnotesize
\setlength{\tabcolsep}{4pt}
\renewcommand{\arraystretch}{0.92}
\begin{tabular}{cccccc}
\hline
\textbf{Task} & \textbf{Acc. (\%)} &
\textbf{Task} & \textbf{Acc. (\%)} &
\textbf{Task} & \textbf{Acc. (\%)}\\
\hline
T1 &98.63 &T5 &98.64 &T9 &100.00\\
T2 &98.64 &T6 &99.46 &T10&100.00\\
T3 &99.16 &T7 &98.40 &T11&100.00\\
T4 &100.00&T8 &100.00&T12&100.00\\
\hline
\multicolumn{5}{r}{$\eta_{\mathrm{task}}$ (\%) :} &
\textbf{100.00}\\
\hline
\end{tabular}
\end{table}

\subsubsection{Processing Latency}

The processing latency of each transmitter-side module is measured frame by frame. Since the dataset consists of several discrete video clips and the object detector incurs a initialization cost at the start of each clip, while the proposed system is intended to run continuously, the first frame of every task is excluded from the statistics. Table~\ref{tab:latency summary} summarizes the mean, median, P95 (95th-percentile), and maximum latency of each stage. The total mean latency is 60.057~ms, which matches the 15~fps camera input.

\begin{figure}[H]
    \centering
    \includegraphics[width=0.75\linewidth]{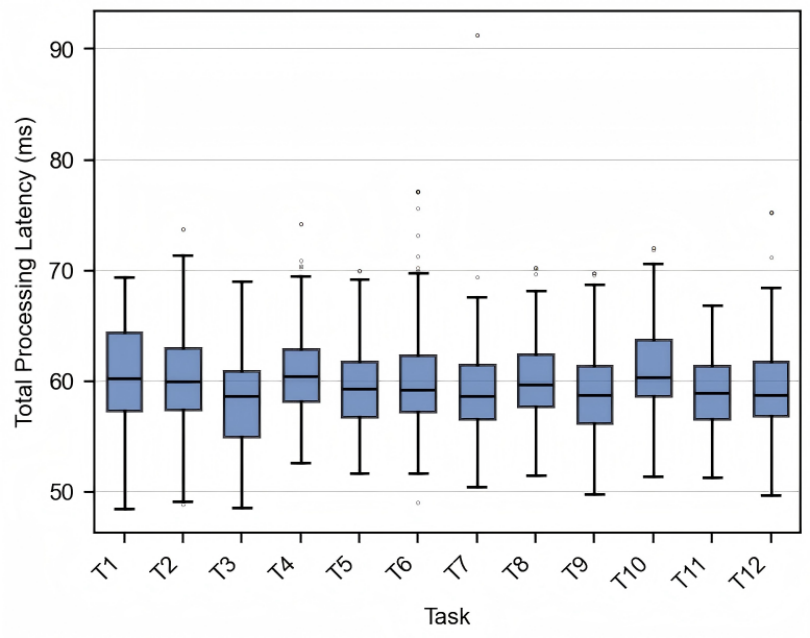}
    \caption{Distribution of total processing latency across all traffic tasks}
    \label{fig:total_processing_latency}
\end{figure}

Fig.~\ref{fig:total_processing_latency} presents the distribution of the total processing latency across all traffic tasks. The latency is tightly concentrated around its median of 59.449~ms, with a P95 of 67.500~ms, and only a few outliers reach the maximum of 136.464~ms, which are mainly caused by detection-model warm-up and system scheduling jitter.

\begin{table*}[!t]
\centering
\caption{Summary of system latency}
\begin{tabular}{ccccc}
\hline
\textbf{Stage} & \textbf{Mean(ms)} & \textbf{Median(ms)} & \textbf{P95(ms)} & \textbf{Max(ms)} \\
\hline
Object Detection  & 50.14 & 49.64 & 56.87 & 126.56 \\
Object Tracking  & 9.84 & 9.73 & 11.53 & 18.52 \\
Semantic Info. Extraction  & 0.06 & 0.00 & 1.00 & 1.62 \\
Semantic Packet Generation  & 0.015 & 0.00 & 0.00 & 1.50 \\
Total  & 60.057 & 59.449 & 67.500 & 136.464 \\
\hline
\end{tabular}
\label{tab:latency summary}
\end{table*}

\subsubsection{Compression Ratio}

\begin{figure}[H]
    \centering    \includegraphics[width=0.6\linewidth]{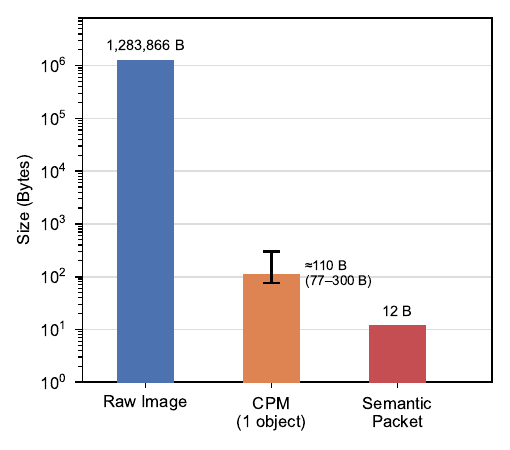}
    \caption{Comparison of Original Image Size and Semantic Packet Size}
    \label{fig:size_comparison}
\end{figure}

To assess the communication efficiency of the proposed framework, the size of the generated semantic packet is compared with that of the corresponding raw image frame and with the estimated size of a Collective Perception Message (CPM) \cite{xhoxhi2023first}. As shown in Fig.~\ref{fig:size_comparison}, each packet retains only perception and intent semantic information of every tracked object, instead of the full image or a CPM packet. Although this paper focuses on single-vehicle scenarios, the packet size is expected to scale with the number of objects and to remain much smaller than raw image transmission. Hence, the proposed goal-oriented architecture delivers decision-relevant traffic semantic information at a very small communication cost, even as the scene becomes more crowded.

\section{Conclusion}

In this study, we proposed a semantic extraction framework for goal-oriented V2X communication to support safe vehicle passage through intersections. The framework formulates intersection-passing scenarios as traffic tasks under Japanese traffic regulations and converts roadside perception results into compact semantic packets containing perception and intent semantic information.

As the transmitter-side core of the envisioned V2X semantic communication system, the framework was implemented and evaluated on CARLA-generated single-vehicle test cases. The results show that it correctly derives the required semantic information for all tested task sequences while satisfying the frame-level real-time requirement of the 15~fps input video. Future work will evaluate transmission and receiver-side decoding
latency and extend the framework to multi-participant scenarios.
It will also investigate message-wise unequal error protection,
building on the layered code construction in~\cite{lu2026layereduep},
with protection requirements determined by the task consequences of message errors.

\printbibliography

@inproceedings{wu2026goal,
author    = {Wu, Wenchao and Wang, Zhe and Deng, Yansha},
title     = {Goal-Oriented Semantic Communication for Joint Vehicle and License Plate Recognition},
booktitle = {IEEE International Conference on Communications (ICC)},
address   = {Glasgow, UK},
month     = jun,
year      = {2026}
}

@inproceedings{li2025robust,
  title={Robust Task-Oriented Communication with Semantic-Aware Masking and Discrete Codebook},
  author={Li, Yundi and Jin, Zhu and Song, Tiecheng and Song, Xiaoqin and Hu, Jing},
  booktitle={2025 IEEE Wireless Communications and Networking Conference (WCNC)},
  pages={1--6},
  year={2025},
  organization={IEEE}
}

@inproceedings{jin2025goal,
  title={Goal-Oriented Communication With Semantic Reconstruction in Vehicular Networks},
  author={Jin, Zhu and Song, Tiecheng and Song, Xiaoqin and Hu, Jing},
  booktitle={IEEE Vehicular Technology Conference (VTC2025)},
 % pages={1--6},
  year={2025},
  %organization={IEEE}
}

@article{bai2024survey,
  author  = {Bai, Zhengwei and Wu, Guoyuan and Barth, Matthew J. and Liu, Yongkang and Sisbot, Emrah Akin and Oguchi, Kentaro and Huang, Zhitong},
  title   = {A Survey and Framework of Cooperative Perception: From Heterogeneous Singleton to Hierarchical Cooperation},
  journal = {IEEE Transactions on Intelligent Transportation Systems},
  volume  = {25},
  number  = {11},
  pages   = {15191--15209},
  month   = nov,
  year    = {2024}
}

@inproceedings{chen2019cooper,
  author    = {Chen, Qi and Tang, Sihai and Yang, Qing and Fu, Song},
  title     = {Cooper: Cooperative Perception for Connected Autonomous Vehicles Based on 3D Point Clouds},
  booktitle = {Proceedings of the IEEE 39th International Conference on Distributed Computing Systems (ICDCS)},
  pages     = {514--524},
  month     = jul,
  year      = {2019}
}

@inproceedings{wang2020v2vnet,
  author    = {Wang, Tsun-Hsuan and Manivasagam, Sivabalan and Liang, Ming and Yang, Bin and Zeng, Wenyuan and Urtasun, Raquel},
  title     = {V2VNet: Vehicle-to-Vehicle Communication for Joint Perception and Prediction},
  booktitle = {Proceedings of the European Conference on Computer Vision (ECCV)},
  pages     = {605--621},
  year      = {2020}
}

@inproceedings{xu2022v2xvit,
  author    = {Xu, Runsheng and Xiang, Hao and Tu, Zhengzhong
               and Xia, Xin and Yang, Ming-Hsuan and Ma, Jiaqi},
  title     = {{V2X-ViT}: Vehicle-to-Everything Cooperative
               Perception with Vision Transformer},
  booktitle = {Proceedings of the European Conference
               on Computer Vision (ECCV)},
  year      = {2022},
  pages     = {107--124},
  publisher = {Springer},
%  doi       = {10.1007/978-3-031-19842-7_7}
}

@article{soga2026latency,
  title={Real-world Latency Analysis of Vehicular Visible Light Communication with Multiple LED Transmitters and an Event-Based Camera},
  author={Soga, Ryota and Shimizu, Tsukasa and Shiba, Shintaro and Kong, Quan and Lu, Shan and Yamazato, Takaya},
  journal={IEEE Vehicular Technology Conference(VTC2026-Spring)},
  year={2026}
}

@INPROCEEDINGS{soga2025,
  author={Soga, Ryota and Shiba, Shintaro and Kong, Quan and Kobori, Norimasa and Shimizu, Tsukasa and Lu, Shan and Yamazato, Takaya},
  booktitle={2025 IEEE Intelligent Vehicles Symposium (IV)}, 
  title={Evaluation of Mobile Environment for Vehicular Visible Light Communication Using Multiple LEDs and Event Cameras}, 
  year={2025},
  volume={},
  number={},
  pages={964-969},
  keywords={Image sensors;Robot vision systems;Bit error rate;Vehicle driving;Receivers;Vision sensors;Light emitting diodes;Cameras;Optical transmitters;Visible light communication},
  %doi={10.1109/IV64158.2025.11097663}
  }

@inproceedings{nakano2026vlccp,
  title={Feasibility Study of VLC-Based Collective Perception for Vehicular Communication},
  author={Nakano, Kosuke and Lu, Shan and Yamazato, Takaya},
  booktitle={IEEE International Conference on Communications},
  %pages={1--6},
  year={2026}
}

@article{yamazato2026ictplatform,
  author       = {Yamazato, Takaya and Nakano, Kosuke and Lu, Shan},
  title        = {Integrating Collective Perception and {ITS} Smart
                  Infrastructure into Regional {ICT} Platforms:
                  Prospects for {V2X} and Visible Light Communication
                  in Autonomous Driving},
  journaltitle = {IEICE Transactions on Communications
                  (Japanese Edition)},
  volume       = {J109-B},
  number       = {2},
  pages        = {98--106},
  date         = {2026-02},
  %doi          = {10.14923/transcomj.2025SII0003},
%  note         = {In Japanese}
}

@online{lu2026layereduep,
  author       = {Lu, Qiming and Lu, Shan and Yamazato, Takaya},
  title        = {Layered Construction of Message-Wise
                  Unequal Error Protection Codes},
  date         = {2026},
  eprint       = {2605.23390},
  eprinttype   = {arxiv},
  eprintclass  = {cs.IT},
  doi          = {10.48550/arXiv.2605.23390}
}

@inproceedings{xhoxhi2023first,
  title={A first study on the spectrum needs for release 2 v2x services},
  author={Xhoxhi, Edmir and Schiegg, Florian Alexander},
  booktitle={IEEE Vehicular Technology Conference (VTC2023-Fall)},
 % pages={1--6},
  year={2023},
%  organization={IEEE}
}

\end{document}